\documentclass[12pt,preprint]{aastex} 

\usepackage{psfig}

\def\runtime{
    \count255=\time\divide\count255 by 60       
    \the\count255:\relax                        
    \multiply\count255 by -60 \advance\count255 by\time 
    \ifnum 10 > \count255 {0}\fi                
    \the\count255                               
    \qquad\the\month/\the\day/\the\year}
\def\runtimeonly{
    \count255=\time\divide\count255 by 60       
    \the\count255:\relax                        
    \multiply\count255 by -60 \advance\count255 by\time 
    \ifnum 10 > \count255 {0}\fi                
    \the\count255}                              
\setbox1\hbox{\today \ at \runtimeonly}
\input colordvi
\def\simge{\mathrel{%
   \rlap{\raise 0.511ex \hbox{$>$}}{\lower 0.511ex \hbox{$\sim$}}}}
\def\simle{\mathrel{
   \rlap{\raise 0.511ex \hbox{$<$}}{\lower 0.511ex \hbox{$\sim$}}}}

\begin{document}
\slugcomment{Accepted by The Astrophysical Journal}            



\title{Timescale Stretch Parameterization  of Type~Ia Supernova $B$-band Light Curves }

\author{G. Goldhaber\altaffilmark{1,2,3}, D.~E.  Groom\altaffilmark{2}, A.~Kim\altaffilmark{2},
G.~ Aldering\altaffilmark{2}, P. Astier\altaffilmark{4}, A.~ Conley\altaffilmark{2}, S.~E.~Deustua\altaffilmark{2}, R. Ellis\altaffilmark{5,6},  S. Fabbro\altaffilmark{4}, A.~S.~Fruchter\altaffilmark{8}, A. Goobar\altaffilmark{7}, I.~Hook\altaffilmark{13},  M. Irwin\altaffilmark{5},  M.~Kim\altaffilmark{2},
R.~A. Knop\altaffilmark{2}, C.~Lidman\altaffilmark{15},
R. McMahon\altaffilmark{5}, P.~E. Nugent\altaffilmark{2},  R. Pain\altaffilmark{4}, N.~Panagia\altaffilmark{8}, 
C.~R.~Pennypacker\altaffilmark{2,9}, S.~Perlmutter\altaffilmark{2,3}, P.~Ruiz-Lapuente\altaffilmark{10}, B. Schaefer\altaffilmark{11,12}, N.~A. Walton\altaffilmark{14}  and T.~York\altaffilmark{2}}

\author{(THE SUPERNOVA COSMOLOGY PROJECT)}

\altaffiltext{1}{e-mail address: {\tt gerson@lbl.gov}}
\altaffiltext{2}{ E. O. Lawrence Berkeley National Laboratory, Berkeley California 94720.}
\altaffiltext{3}{Center for Particle Astrophysics, U.C. Berkeley, California 94720.}
\altaffiltext{4}{LPNHE, CNRS-IN2P3, University of Paris VI \& VII, Paris, France.}
\altaffiltext{5}{Institute of Astronomy, Cambridge University, Cambridge, U.K.}
\altaffiltext{6}{Current address: California Institute of Technology Pasadena, California.}
\altaffiltext{7}{Fysikum, University of Stockholm, S-11385 Stockholm, Sweden.}
\altaffiltext{8}{ Space Telescope Science Institute, Baltimore, Maryland.}
\altaffiltext{9}{Also Space Sciences Laboratory, U.C. Berkeley, California.}
\altaffiltext{10}{Department of Astronomy, University of Barcelona, Barcelona, Spain.}
\altaffiltext{11}{Department of Astronomy, Yale University, New Haven, Connecticut.} 
\altaffiltext{12}{Current address: University of Texas, Austin, Texas.}
\altaffiltext{13}{ Institute for Astronomy, The University of Edinburgh, 
     Edinburgh EH9 3HJ, Scotland.}
\altaffiltext{14}{Isaac Newton Group, La Palma, Spain.}
\altaffiltext{15}{European Southern Observatory, La Silla, Chile }

\begin{abstract}
$R$-band intensity measurements along the light curve 
of 
Type Ia supernovae discovered by the Supernova Cosmology Project (SCP) are
fitted in brightness to templates allowing a free parameter the time-axis width factor  $w\equiv s \times(1+z)$.
The data points are then individually aligned in the time-axis, normalized and $K$-corrected back to the rest
frame, after which the nearly 1300 normalized intensity measurements
are found to lie on a well-determined common rest-frame $B$-band curve which we call the ``composite curve''.
The same procedure is applied to 18 low-redshift Cal\'{a}n/Tololo SNe with $z < 0.11$; these nearly 300  $B$-band photometry points are found to lie on 
the composite curve equally well.
The SCP search technique produces several measurements before
maximum light for each supernova.
We demonstrate that the linear stretch factor, $s$,
which parameterizes the light-curve timescale appears independent of $z$,
and applies equally well to the declining and rising parts of the
light curve.  
In fact, the $B$ band template that best fits this composite curve
fits the individual supernova photometry
data when stretched by a factor $s$ with  $\chi^2/DoF$  $\approx$ 1, thus as well as any 
parameterization can, given the current data sets.
The measurement of the date of explosion, however, is model 
dependent and not tightly constrained by the current data.
We also demonstrate the 
$1+z$ light-curve time-axis broadening expected from cosmological expansion. This argues strongly against alternative explanations, such as tired light, for the redshift of distant objects.
\end{abstract}
\keywords{supernovae: general -- cosmology: observations}

\section{Introduction}
\label{sect:intro}

In systematic searches over the last 12 years the Supernova Cosmology
Project (SCP) has discovered and studied over 100 supernovae, most of which
have been spectroscopically identified as Type~Ia supernovae
(SNe~Ia) at redshifts up to $z = 1.20$.  
Forty-two of these events have so far been used for measurements of 
the cosmological parameters $\Omega_M$ and
$\Omega_\Lambda$ (\citet{perl99}, 
hereafter ``P99''; see also Perlmutter et al. 1995 and 1997, 
hereafter ``P95'' and ``P97''). See also Riess et al. 1998 who used  sixteen high redshift SNe in a parallel study of the cosmological parameters.
An important element of these studies 
has been the recognition of homogeneity within a one parameter family
of SN Ia light curves.  

Several approaches have been used to characterize the SN Ia 
light-curve family.
\citet{phillips93} first suggested that the range of SNe Ia light curves
might be grouped into a single family of curves parameterized by their
initial rate of decline (see also the earlier suggestions of 
\citet{pskovskii77,pskovskii84}).
\cite{phillips93}
and \cite{hametal95} further observed that the absolute magnitudes of SNe Ia at maximum light were tightly correlated with this decline rate in
the $B$-band light curve --- brighter SNe Ia have wider, 
slower declining light curves. 
Riess, Press, \& Kirshner (1995, 1996) introduced an alternative
characterization of this range of light curves, in which a scaled
``correction template'' is added or subtracted from a standard
template, effectively widening or narrowing the shoulders of the
light curve  and thus varying its timescale.  In our analyses (P95, P97, P99)
we instead parameterized the timescale of the SN Ia light curve by 
a simple
``stretch factor'', $s$, which linearly stretches or contracts the time
axis of a template light curve around the date of maximum light, and 
thus affects both the rising and declining part of the light curve for each
supernova.  
All three methods of characterizing the light curve shape
 or timescale have been used to calibrate the peak magnitudes.

In this
paper, we use the dataset of supernovae studied in P99 to 
examine the empirical behavior of SN Ia light curves,
and show that this single parameter, the stretch factor $s$, can 
effectively describe almost all of the diverse range of $B$-band 
light curve shapes over the peak weeks.
The stretch factor method was originally introduced and tested (P95)
without the benefit of either detailed or early data.  We can
now test whether it applies to the entire curve, rather than just the
post-maximum  decline and determine the scatter about this scaled curve.



This paper is organized as follows.  In Sect.~\ref{sect:data}
we present the set of supernova and the data used in this analysis.
In Sect.~\ref{sect:standard} we use fits to a pre-existing template light curve
to cross-compare our data from each supernova
and to produce a single composite plot of all the data together.
We use the dataset of the composite plot to test the validity of our light-curve
model and of the stretch parameter. 
In Sect.~\ref{sect:dilation}, we
test the hypothesis that redshift is due to cosmological 
expansion by studying
the predicted $(1 + z)$ broadening of the light curves and we look for
evidence for possible supernova evolution.  
Finally in  Sect. 5, we present a general discussion. 
\section{Data}
\label{sect:data}

A paper based on our first seven SNe gives the details of our
photometric and spectroscopic measurements, as well as the analysis
methods (P97; see also P99). 
Most of the objects were followed photo-metrically for over
a year.  Except for a few SNe discovered before 1995, most spectra 
were obtained at an early enough epoch at the Keck 10-m telescopes to observe 
both the characteristic Type~Ia SN spectrum and the host galaxy spectrum.
Redshifts were determined primarily from narrow emission and absorption lines in the host galaxy spectra.
In this paper we are mainly concerned with 35 of the 42 fully-analyzed SNe
reported in P99 which are  listed in Table~\ref{scpdata}.
 The redshifts for this 35-supernova subset,
$0.30 \le z \le 0.70$, are such that ``cross-filter $K$ corrections'' are 
made from the observed Kron-Cousins 
$R$ to the rest-frame $B$ band \citep{kim_kcorr96, nuge_kcorr99}, 
for which lower-redshift
data are available for comparison.  
Of the 42 SNe, 7 are omitted here as follows: Two $z < 0.3 $ and
three $z > 0.7 $ supernovae, which have to be  $K$-corrected to
the $V$ and $U$ bands respectively, are not included here.  One distinctly faint
outlier, SN1997O, about 1.5 magnitudes fainter than the average, is also omitted. We have checked that it makes no significant difference whether this SN is included or not. SN1994G is omitted because of sparse
and late $R$ photometry; it was observed primarily in the $I$-band.
We consider one of the supernovae included in this study, SN1994H, 
to be unlikely to 
be Type Ia, based on our more recent analyses (P99, and \citet{nuge_kcorr99}).
It was included here for cross-comparison with the analyses of P97 and P99
(in which it was excluded from the primary analyses);
we have checked that its light curve mimics a Type Ia SN quite well, and that including it in this study makes no significant difference. The lower-redshift sample used here consists of 
18 SNe selected from the 29 SNe of the
Cal\'{a}n/Tololo set \citep{hametal95,hametal96}, based on the criterion that they were discovered prior to 5 days after
maximum light. 
 This homogeneous sample is chosen since it is the low redshift sample used in P99 which required SNe in the Hubble flow ( beyond $z \approx 0.02$) for comparison with the high redshift sample. We have also examined an additional set of 18 SNe which satisfy the criterion of discovery prior to 5 days after maximum, as discussed below.

\begin{table}
\dummytable\label{scpdata}
\end{table}

\begin{table}\dummytable\label{hamuydata}
\end{table}

\begin{table}
\dummytable\label{template}
\end{table}

The SCP search strategy (P95) ensures early points by obtaining  a set of several 
baseline ``reference images'' a few days after new moon, and another set of several ``search
images'' about three weeks later.  This ensures almost two weeks of
dark time after discovery, for spectroscopic and photometric
follow-up. Supernova candidates are found by differencing the pairs of images.  The supernova may or may not be present, but in any case is much
fainter on the reference set; this is established from archival
references or references obtained a year later. The procedure yields one, or even two, image sets for most SNe on the early rising light curve.

\section{Reduction to a Standard $B$-band Light Curve}
\label{sect:standard}

In the data analysis as described in P97 and P99, each SN's flux light-curve
points are
fitted to a $R$-band template light curve using the nonlinear fitting
program MINUIT \citep{MINUIT}. For $R$-band photometric data the
function
\begin{equation}
I(t)/I_{max} = f_R\Big((t-t_{\rm max})/s(1+z)\Big) + b
\end{equation}
is fitted to the data by adjusting the intensity $I_{max}$ at maximum light, the time of maximum $t_{\rm max}$, the stretch factor $s$, so that
the width factor $w\equiv s(1+z)$ scales the template time axis, and 
a baseline level, $ b$ whose amplitude is found to be $<  0.02$ and allows for small corrections in the background galaxy subtraction.  

The function $f_R$ is generated from a $B$-band template (in magnitude)
($F_B(t)$) $K$-corrected to the $R$-band for the given redshift, $z$, converted to flux and 
renormalized to unity at light maximum: $f_R(0)=1$.  We use as a template
``SCP1997'', the same used in P97 and P99, and given in flux in Table 3  \citep{leibundgutthesis}.
SPC1997 was based on empirical fits
by Leibundgut (shown in column~2 of Table~\ref{template} in flux ) , tabulated for 
$-5\leq t \leq 110$~days \cite{leibundgutthesis},
and then extended by us with updated data to both late and early
times.
The
constant slope in magnitude $dm_B/dt = 0.0166$/day for days 50--110 
was extrapolated to even later times.
A piecewise quadratic curve was adjusted
to give a reasonable fit to the data for 8 nearby Type~Ia
SNe \footnote{SN1990N, SN1990af, SN1991T, SN1992A, SN1992bc, SN1992bo, SN1994D and SN1994ae} for $-15 \leq t \leq -5$~days, and the slope defined by the two
earliest points in magnitude ($-0.5041$/day) was extrapolated to earlier times. 
This linear extrapolation in magnitude is exponential in flux. Only
a handful of measurements were available for $t<-10$~days, and the
earliest behavior rested on single measurements of SN1994ae at
$-12.9$~days and SN1994D at $-14.8$~days in the observer system.
This template was used for our
cosmology analysis (P97, P99).
Note that since all fits to data were performed starting with this
template, this template has a stretch factor of $s = 1$ by definition.

The results of the fits are shown in
Table~\ref{scpdata} which
gives the redshift, $z$, the measured width factor, $w$, in
the observer system and the corresponding stretch factor, $s$, as well as the
the measurement errors, $\sigma_w$ and $\sigma_s$, and the $\chi^2/DoF$ for the MINUIT fits.
 Table~\ref{hamuydata} gives the same information for the
Cal\'{a}n/Tololo low-redshift data.

 The systematic uncertainty in the time-dependent
$K$ corrections, recalculated for P99, are estimated as $ < 0.02 $
magnitudes for the light-curve phases later than $\simeq-14 $ days.  
An effect we need to consider in determining the early time
light-curve behavior is how uncertainties in the $K$-corrections at
these early epochs propagate into uncertainties on the $K$-corrected
$B$-band fluxes. In particular, we need to examine all the points
which occur prior to day $-14 $ with respect to maximum light. Later than
this epoch we have the necessary spectroscopy of a variety of SNe~Ia
to perform accurate $K$-corrections. Prior to this we have tied
all the $K$-corrections to our knowledge of the spectral energy distribution (SED) of SNe~Ia at this earliest epoch.

At a redshift of $z \approx 0.5$, the uncertainty in the
$K$-corrections as a function of epoch is negligible due to the nice
alignment of the $B$- and $R$-band filters \citep{kim_kcorr96}. For SNe~Ia at redshifts
below this value (to $z=0.3$, our nominal cutoff for SN~Ia included in
this study) the $B$-to-$R$ $K$-correction ($K_{BR}$) involves
knowledge of SED redward of the $B$-band filter. For SNe~Ia beyond
this redshift (to $z=0.7$, our other cutoff) $K_{BR}$ relies on our
understanding of the $U$-band behavior of SNe~Ia. For a simple and
very conservative test, we can gain an understanding of the
uncertainties by looking at the $K$-corrections at these two extreme
redshifts, which involve the greatest extrapolation and hence would
produce the largest error.

For $z=0.3$ and day $ -14 $ we find  $K_{BR} = 0.51$ for a SN~Ia from a spectral analysis. This value corresponds to the  value that a $\approx$ 15,000 K blackbody would give
 at this redshift. From the models of \citet{hoflich95},
\citet{nuge_hyd97} and \citet{lentz99}, we can see that in the optical spectra
redward of 4000\AA, at the earliest epochs after explosion, SNe~Ia
closely approximate blackbodies with a very minimal
contribution from the lines ($< 10\%$ ) to the overall SED. The exact temperature is not critical in going from 15,000 K to 100,000 K. Using a 100,000 K
blackbody as a model of the earliest epoch for a SN~Ia, (a gross
overestimate of its temperature since the $\gamma$-rays from the decay
of $^{56}$Ni have not had enough time to propagate out and heat up the
atmosphere) $K_{BR}$ only changes by 0.07 magnitudes This is well below the
the difference that can be distinguished by the data sets at these very early epochs.
Only one SN~Ia above a redshift of 0.6 (where the $K$-corrections
start to involve a decent amount of extrapolation), SN1995at,
contributes to the early light curve prior to day $-14$. SN1995at was
observed at day $ -16.3$ and was at  $z=0.655$. The effect of this
single photometry point on our calculations is negligible.
In addition, all of the models mentioned above, show very little
difference in the overall SED over the epochs $-14$ to $-16.3$ days.

\subsection{A Composite Light-curve}

We now introduce the concept of a ``composite light curve'', constructed
by linearly compressing or expanding 
 the time axis for each SN such that all the low-$z$ and high-$z$ data 
points can be plotted on a single curve. To study the light-curve 
data in this composite form, a further step is added to
the procedure of P97 and P99: 
After the time of
maximum  $t_{max}$ and the stretch factor for each SN is obtained from the fit,
the individual $R$-band data
points are $K$-corrected back to points in the equivalent rest-frame $B$ band.

\begin{figure}
\plotone{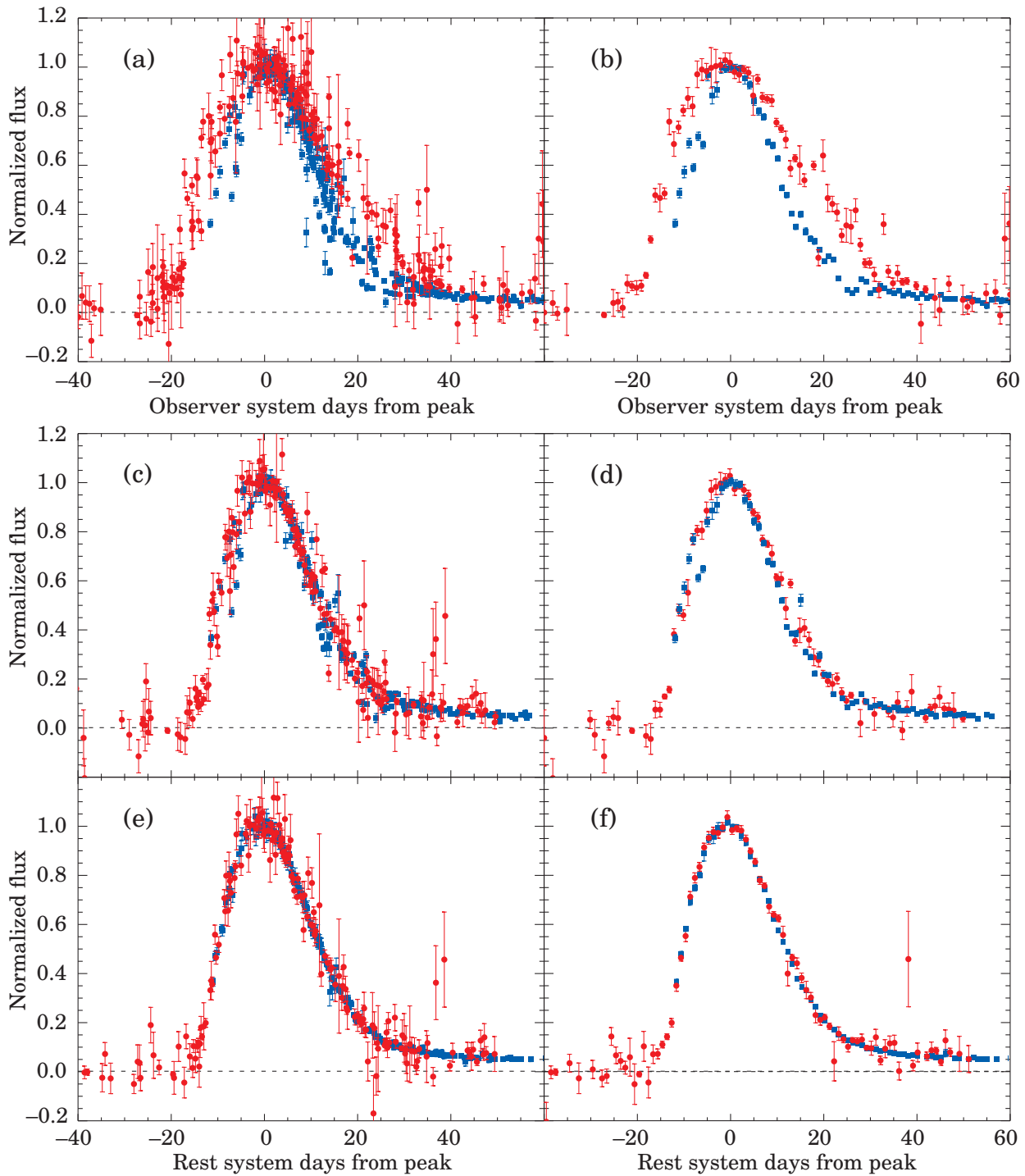}
\caption{\looseness=-2(a) The photometry points for the 35 SCP (full red circles) and 18
Cal\'{a}n/Tololo SNe (blue squares), fitted to Parab-18 with the maximum flux normalized to unity 
and the time of maximum adjusted to zero
in the observer system.    (b) shows the same data as in (a)
averaged over one-day intervals and over each set of SNe. 
(c) and (d) show the same data, transformed to the rest system. In (e) and
(f) the time axis for each photometry point is additionally divided by the corresponding stretch factor $s$.
\label{light curves}
}
\end{figure}

Figure~\ref{light curves} presents all photometry points in composite
light curves for the template ``Parab-18'' which is an improved template over SCP1997, as discussed below. 
The left-hand
figures (a), (c), and (e) show a data point for each night's observation
of each supernova, whereas the right-hand figures (b), (d), and (f) show
one-day averages over all supernovae.
The K-corrected SCP points for SNe with $0.30 \le  z \le 
0.70$ are shown as solid red circles. Shown as blue squares are
$K$-corrected points ($K_{BB}$) for the 18 SNe from the Cal\'{a}n/Tololo study in
the $B$-band (\cite{hametal96}) with $z < 0.11$. Figures~\ref{light curves}
(a) and (b) show the $B$-band photometry points in the observer system 
displaced
to $t = 0$ at light maximum, and normalized to unit intensity at $t = 0$.
There are no
corrections for stretch or width; times of observation relative to
maximum light are used.  
In
Figures~\ref{light curves}(c) and (d) we have transformed the time axis
from the observer frame to the rest frame by dividing by the
appropriate $1+z$ for each data point of each of the SNe.  Finally, in
Figures~\ref{light curves}(e) and (f) for each supernova we also divide
the time scale for each point by the fitted stretch
factor~$s$. By this stage essentially all of the dispersion has been removed, and
the corrected points for both the SCP and Cal\'{a}n/Tololo data fall on a common curve at the level of the 
measurement uncertainty,
of typically $2-4\%$.
It is remarkable that application of this single stretch timescale parameter 
results in such a homogeneous composite curve.
  
\subsection{The Templates}

While we used the exponential rise in our fits to $s$, $t_{max},$  
$I_{max}$ and $b$ with the SCP1997 light-curve template,
in reality there must be a definite explosion date $t_{exp}$. To find an improved early template for the composite light curve we use
 a parabolic approximation to the initial light curve,
and determine its two parameters, explosion date $t_{exp}$ and timescale
parameter $a$, by calculating a least-squares fit  
of the function $I(t)/I(0)=a(t-t_{exp})^2$
to the UN-averaged stretch-corrected composite flux data (which, unlike magnitudes, have Gaussian errors).
(Since the fits were to the stretch-corrected
composite data, the actual explosion day of any individual 
supernova is $ t_0 = s \times t_{exp}$.)
We determined that fitting the early-time parabola to the 
later-time exponential template 
(matching both the value and first derivative) was not very sensitive to the join date in the epoch around -10 days. 

Using this additional form for the early part of the light curve, we can 
compare the combined composite light curve of
Figures~\ref{light curves}(f) with
three forms of the template $F_B(t)$:

\begin{enumerate}
\item
The SCP1997 template:  This light-curve template (listed in Table 3 ) begins with a linear rise in 
magnitude (exponential in flux); it was used in P97 and P99. The width factor 
and stretch factor values given in Tables 1 and 2 come from fits to this 
template and are supplementary to Tables 1 and 2 in P99 . 
\item
The ``Parab-18'' template: This light-curve template (listed in Table 3 ) begins with a parabolic rise in flux; its explosion date at $-17.6$ days is based  on a parabolic fit to the composite data (starting from the SCP1997 template).  A cubic spline fit to the data beyond the ``join date'' at $-10$ days improves the template fit as compared to SCP1997 (or Leibundgut), in particular near -5 days.  This template thus gives the best fit to the SCP data, although the fit to Parab-20 (see next item) is nearly as good. Whenever so stated the analyses in this paper are based on this template.
\item
The ``Parab-20'' template:  This light-curve template (listed in Table 3) also begins with a 
parabolic rise in flux, but with an explosion date at $-20.0$ days based 
on the recent early data of Riess et al. (1999a).  
At later times, after day $-10$, it is the same as Parab-18, and thus differs
by only a small amount from the SCP1997 template.
This template also gives an acceptable fit to the composite data, as
also found by  Aldering, Knop, and Nugent (2000; hereafter AKN2000).
\end{enumerate}

\begin{figure} 
\centerline{
\psfig{figure=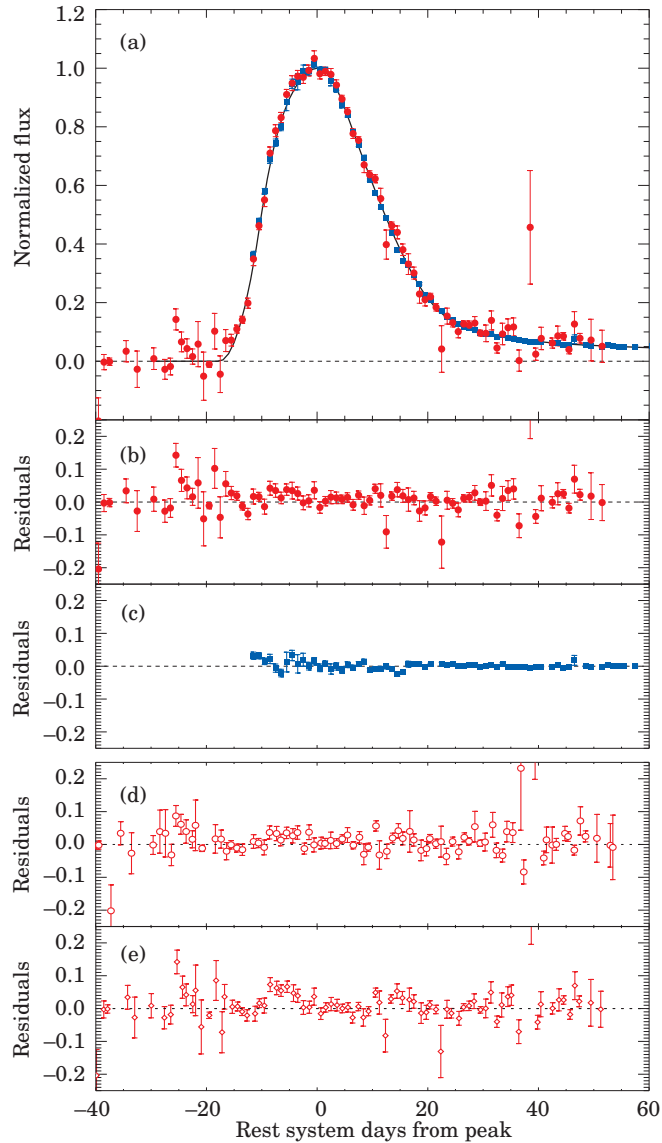,height= 6.0in}}
\caption{ (a) The same distribution as in Figure 1(f) with the SNe in the two data sets fitted to Parab-18. The template curve for Parab-18 is shown in black. (b) Residuals for the SCP (solid red circles) photometry fits to Parab-18.  Here the epoch between -40 and -20 days 
has no SN signal, the data comes from the reference images which have shorter exposures and hence larger errors (see Table 4). (c) Residuals for the Cal\'{a}n/Tololo (blue squares) photometry fits to Parab-18 (see Table 6                    ). (d) The open red circles show the  residuals for the SCP data from the fit to Parab-20 (template not shown), where the early rise is based on the data of Riess et al. (1999a).  See Table 5. (e) Residuals for the SCP data  (open red diamonds) fitted to the template SCP1997  (template not shown). Note that the ``bumps" near -5 days and 15 days (see also Table 7)  were eliminated for the templates Parab-18 and Parab-20. }
\label{template_fit} 
 \end{figure}

Figure~\ref{template_fit} 
a based on Parab-18, shows that this template
  accounts well for
all the data points in our composite sample. Both the SCP and the  Cal\'{a}n/Tololo data scatter about this template
with a dispersion that is expected from their measurement uncertainties. 
From the residual plots of Figure~\ref{template_fit} (b) and (d)
it is clear that this day-by-day uncertainty is less than $2-4\%$
(0.02--0.04 mag) for almost the entire period from day $-10$ to day 40. The $\chi^2$ distributions for the residuals is given in Table 4 (Parab-18) for the 35 SCP SNe in 5 day intervals from -25 to 40 days and, in the last line, for this entire interval. 

A good fit to the data is also obtained for Parab-20, corresponding to $t_{exp} = -20$ days, for which we give the residual plot in Figure~\ref{template_fit}(d) and the
$\chi^2$ distributions for the residuals in 
Table 5.  Fitting the data to this new template, Parab-20, results in
the fitting program MINUIT slightly re-adjusting  stretch $s$, 
intensity at maximum light $ I_{max} $ and the time of 
maximum $t_{max}$ for each SN.  The average change
in $m_B^{effective}$ from the values found using the SCP1997 template, for the 35 SCP SNe, is $-0.005$. The corresponding average change in $m_B^{effective}$ for the 18 Cal\'{a}n/Tololo SNe is $-0.014$. Thus the net change in difference between low and high redshift  $m_B^{effective}$ is 0.009. This is the only quantity that enters into the cosmological measurements via the distance modulus.
 The rising part of the curve for 
both templates Parab-18 and Parab-20  fits our data  equally well with 
only a negligible  change in the overall $\chi^2/DoF$ 
from 0.905 to 0.906 for the epoch -25 to 40 days.
Either of the above templates are an improvement 
over SCP1997 for which we obtain $\chi^2/DoF = 0.966$ (see Table 7). 
However within the accuracy of our data, any of the 3 templates 
is perfectly acceptable. \footnote {Given the low-redshift data presented by Riess et al (1999a), we believe that 
the Parab-20 template is likely to be the best current estimate for
a single common stretch-parameterized light curve.}
Note that the composite light curve depends in fine detail on the template used to fit and then match the peak and stretch of each individual SN's light curve.

 For the Parab-18 template we found $ a=0.0085 \pm 0.0010 $ and $t_{exp} = -17.6 \pm 0.5$ days; a preliminary value of  $ -17.5 \pm0.4$ was presented in the conference reports of \cite{groomaas} and Goldhaber (1998a,1998b).
However it is important to reckognize that these uncertainties reflected only 
 the error in mapping the fixed composite light curve
to a parabola after fitting with an assumed template (and thus did not include errors which depend on the shape of the early portion of the template used).
Since the fits are sensitive to the early portion of the light curve, the total uncertainty on $t_{exp}$ is much larger. 

 For comparison with AKN2000 we have analyzed the same more restrictive set of 30 SCP SNe with $0.35 \le z \le 0.65$ and leaving out SN1997aj. \footnote {The fits for SN1997aj undergoe a large change  in stretch, from 0.94 to 1.52, in going from the template SCP1997 to Parab-20. Because of this feature this SN was excluded by AKN2000 and in the 30 SN sample here
 However in the $\chi^2 $ distributions quoted in Tables 4 and 5 all 35 SNe were included.} 
A simple $\chi^2$ study of these SNe, fitting them to a series of parabolic-rise templates as a function of explosion dates, yields higher statistical uncertainties   $t_{exp} = -( 17.8  _{-1.0}^{+1.9} ~statistical)$ corresponding to a 1$\sigma$ range of $-19.7$ to $-16.8$ days. 
This statistical uncertainty is consistent with the more comprehensive analysis of AKN2000, which accounted for the
full multidimensional probability distribution as well as upper limits on several systematic uncertainties, and found   $t_{exp} = -( 18.3  _{-1.2}^{+1.2} ~statistical ~_{-1.9}^{+3.6} ~systematic)$ at $t_{join} = -10$ days. Note that the exact
shape of the early portion of the light curve does not significantly affect the
values or uncertainties of the cosmology results of P99 (see AKN2000 as well as our results above).

 Only future data will be able to tell whether all type Ia SNe can be adequately described by $s$ only or whether some variation in $t_{exp}$ will also be needed. This will require very early SN detection, corresponding K-corrections, and spectra as is envisioned for the satellite project {\em SNAP}, currently under design. \footnote{See http://snap.lbl.gov for information on SNAP, the SuperNova Acceleration Probe.}

Table 6
gives the corresponding data for the 18 Cal\'{a}n/Tololo SNe from -10 to 40 days. Since there is no significant data before -10 days, 
Parab-18 and Parab-20 coincide in this case.
For these data the first two epochs give a rather high $\chi^2/DoF$ of $\sim2$. We note that at these two epochs the data come primarily from only two SNe; of these only SN1992bc has a poor overall $\chi^2$ ( see Table 2) and in particular is responsible for the high  $\chi^2$ values at the two epochs. 
To improve the understanding of the early epochs  we have also studied 18 additional low redshift SNe which are however not discovered in a  homogeneous search and are not all in the Hubble flow.
These SNe consist of 16 of the 22 SNe given by \cite{riess_22}, which fulfill the criterion of being discovered prior to 5 days after maximum light, as well as SN1990N from \cite{lira98} and SN1998bu from \cite{suntz98} as extended to earlier times (before $-10$ days ) by Riess et al (1999a). This low redshift sample fits the Parab-20 template from -15 to 40 days with a $\chi^2/DoF$ of 1.015, thus agreeing well with the  18 Cal\'{a}n/Tololo SNe in the overlap region. This study of the second set of 18 SNe will be presented in a future publication.

\subsection{The Universality of the Stretch Factor}
 
  The stretch factor applies both to the rising and to the falling part of the  light curve. A qualitative impression of this can be obtained from a comparison of  Figures 1(c) and (e) as well as (d) and (f). To obtain a  quantitative result we took two approaches. In Table 8  we show the $\chi^2$ distribution for the case in which only $1 + z $ but not $s$ has been factored out of the observer frame distribution, this corresponds to Figures 1 (c) and (d). Here we note large $\chi^2$ values both before and after maximum light, indicating that ignoring $s$ affects both distributions. Secondly we have taken the individual photometry points, of the composite curve and split them into two groups at epoch $t = 0$. We then fitted each half with MINUIT to determine a stretch.
This gave $s = 1.03 \pm0.04$ with $\chi^2 = 352$ for 366 DoF for the epoch $< t_{max}$ and  $s = 1.00 \pm0.02$ with $\chi^2 = 894$ for 996 DoF for the epoch $> t_{max}$. The period $> t_{max}$ encompassed all the photometry points out to over 1 year. The equality of $s$, within statistical errors, for the two groups shows that the same value of $s$ applies to both.

 As can be noted from Tables 4 and 6 as well as Figure 2 (b) and (c) the factor 
$s$ combined with the factor $1 + z $ brings all the 35 SCP SNe and the 18 
Cal\'{a}n/Tololo SNe in agreement with a single curve from -25 to over 25 days 
as well as can possibly be determined  with the present datasets.  
The stretch factor parameterization appears to work quite well at least out to day 40, although the high-redshift data is less constraining at these late dates, and there is evidence of at least two low-redshift supernova that may vary from the template at these later epochs (see AKN2000).

\section{Test of Cosmological Light Curve Broadening and Check for SN Evolution}
\label{sect:dilation}
 
These data also produce compelling evidence that the observed explosion
of the supernova itself is slowed \footnote {This light curve broadening has sometimes been
described as ``time dilation'', at variance with usage in physics texts
({\em e.g.}~\cite{weinberg72}).} by the factor $1+z$. 
This provides independent evidence for cosmological expansion as the
explanation of redshifts.  Although this hypothesis has proven to be
consistent with observation for over half a century, persistent doubts
are still occasionally expressed ({\it e.g.}
\cite{maric77,chow81,laviolette86, arp87, arp90, arp94, na93}).
Surprisingly, until recently very few direct tests of this expansion
have been performed. A test by \cite{SP91}, who calculated the surface
brightness of brightest cluster galaxies over a range of redshifts, showed
compelling evidence for expansion but did
not reach a definitive conclusion due to possible systematic errors.
An argument has also been made that
some gamma ray bursts (GRBs) should be at cosmological distances
because of the observation that for the ``longer GRBs'' the length of
the bursts was inversely correlated to the brightness of the GRBs
\citep{piran92,norris95}.  The discovery of  GRBs at
cosmological distances strengthens this argument (see for example
\citet{metzger97}); however since the intrinsic length of a given GRB is
unknown, this remains a qualitative argument \citep{lee2000}.

Using supernova light curves to test the cosmological expansion was
first suggested by \cite{wilson39} (see also \cite{R74}).  Over the
last decade it has become clear that Type~Ia SNe, found nearby and at
cosmological distances, provide superb and precise clocks for such
tests.  We presented the first clear observation of the $1+z$
light curve broadening, based on our first seven high $z$ Type~Ia SNe
\citep{goldhaberaigua}.
\cite{leibundgutdilation} later also gave  evidence for the
effect, using a single  high-$z$ supernova. More recently,
\cite{riess_age97} showed evidence that the spectral
features of Type~Ia SNe can be timed sufficiently well to measure the
time interval between two spectra taken 10 days apart in the observer
system. Applying this method to one supernova gave results consistent
with $1 + z$ light-curve broadening at the 96.4\% confidence level.
With the current dataset, we can now demonstrate the light-curve broadening
with a larger, statistically significant sample.

\begin{figure}
\centerline{
\psfig{figure=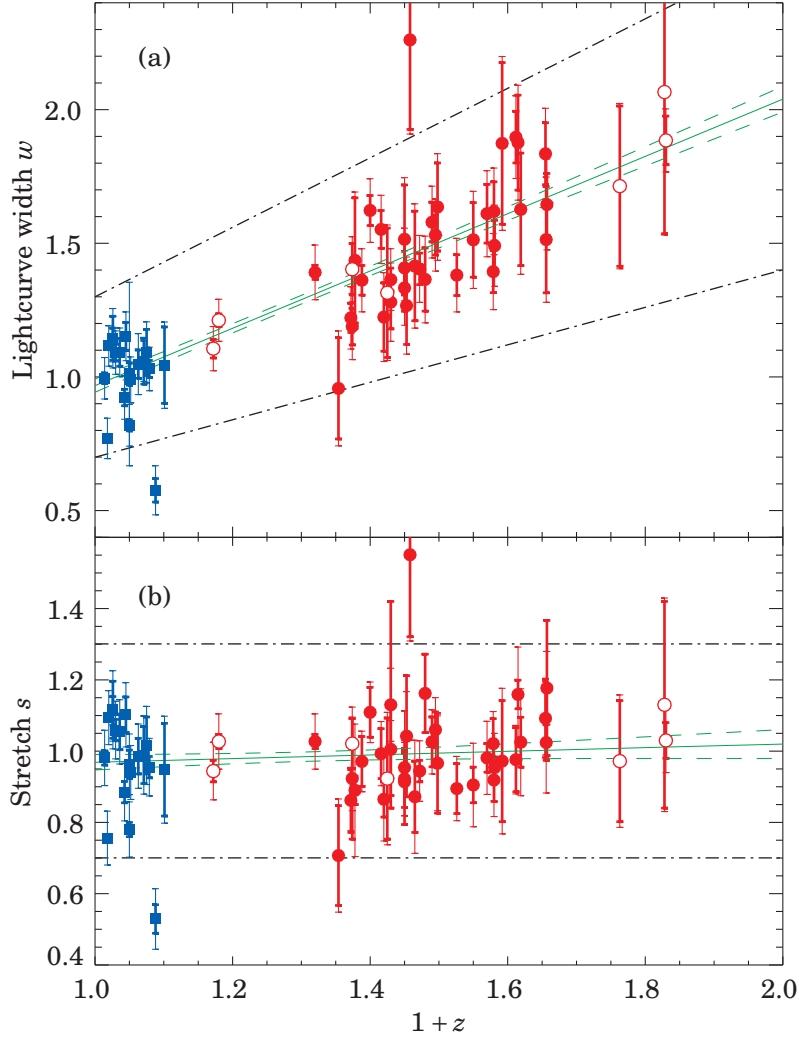}}
\caption{(a) The observed light-curve width factor $w$ vs $1+z$. The blue squares correspond to the Cal\'{a}n/Tololo SNe, the solid red circles are for the subset of 35 SCP SNe used in this paper. The open red circles are for the
remainder of the 42 SCP fully-analyzed SNe. The band
delineated by the black dash dotted lines corresponds to stretch values 0.7 to
1.3  which encompass the bulk of the data, except for two outliers. The green line is the best linear fit to the data as discussed in the text. The band delineated by the two green dashed curves correspond to  the $ \pm 1\sigma $ values. (b)  The stretch $s$ vs $1+z$. Stretch is defined as the observed
light-curve width $w$ divided by $1+z$ for each SN. The points and lines
are defined as in (a).}
\label{width_1plusz}
\end{figure}

Light curve width factor $w$ and stretch factor $s$ versus  $1 + z$ are 
shown in Figure~\ref{width_1plusz}(a) and Figure~\ref{width_1plusz}(b) 
respectively for the Cal\'{a}n/Tololo and SCP SNe.  We are interested in the 
possible variation of $w$ and $s$ with $z$. We test for the $z$ dependence of 
these distributions by fitting  straight lines to the data. If the SNe in the 
distant past were different, i.e. due to
evolution effects, the distribution might show a slope $ds/dz$.  The fits
described below were carried out for the entire 18 Cal\'{a}n/Tololo SNe
together with 40 of the SCP SNe.  Two SCP SNe, SN1992bi and 
SN1992br, have stretch factors
outside the range $0.7 < s < 1.3$, and 
are excluded from the fits as outliers representing 
non-Gaussian tails to the distributions or aberrant objects. 
As discussed in P99, these will not be  important for the present analysis.

In performing the fit of the
function $s=a+b\,z$ to the data shown in Figures~\ref{width_1plusz}(b) the
total uncertainty
for each point is  $\sigma_{s'} = \sqrt{\sigma_s^2 + \delta_s^2}$
where $\delta_s^2$ is the intrinsic stretch dispersion
and $\sigma_s$ from Tables 1 and 2  is from the individual light-curve fit
uncertainty. We estimate 
this intrinsic dispersion by requiring that the reduced $\chi^2$ is near unity. This yields  an rms deviation $\delta_s\approx0.08$.
 The intrinsic width $\delta_s$
of the stretch factor distribution can also be obtained from a Gaussian fit to Figure 4 in P99.
This  gives a consistent value for  $\delta_s$ of $ \sim 0.1$.
 The errors used in the fit to the $w$ distribution shown in Figures~\ref{width_1plusz}(a) are treated similarly.
In Figure~\ref{width_1plusz} (a) and (b) both errors $\sigma_s$ and  $\sigma_{s'}$ are shown on the error bars.

We find that our data is consistent with $1 + z $ lightcurve broadening and a redshift independent SN stretch distribution.
Fits to Figure~\ref{width_1plusz} (a) and (b) yield different information.  The linear fit shown in Figure~\ref{width_1plusz}(a) has $dw/dz =
1.07\pm0.06$. If there were no $z$ dependence in $w$ this slope 
would be $0$, here assuming no evolutionary change in $s$. Hence the evidence for the presence of a $1 + z$ 
factor is  $1.07/0.06 \approx 18$ standard deviations.
Figure~\ref{width_1plusz}(b) shows a slope $ds/dz =
0.05\pm0.05$. The extent by which this slope differs from $0$ 
measures the possible evolution effects on $s$, here assuming a $1 + z $ dependence of $w$. The result indicates 
that $ds/dz < 0.09$ at the $95\% $ confidence level out 
to $z \sim 0.8$. We obtain essentially the same results 
from fits in which the 2 outliers are included (for a total of 60 SNe), 
and when the 35 rather than the 42 SCP SNe, less 2 outliers, are 
used (for a total of 51 SNe). Note that this analysis does not account for sample selection effects --- such as possible preferred discovery of high stretch SNe near the flux limit of each SNe search --- which may need to be accounted for in future datasets.


When we compare this result to the alternate theories, it is
clear that they are severely challenged or simply ruled out.
The tired light theories (\cite{zwicky29,hubble_tolman35,hubble36,
maric77,chow81,laviolette86} would not yield this slowing
of the light curves, and thus do not fit this dataset.  Variable
mass theories  (\cite{narlikar77,na93,na97} would apparently require a series
of coincidences:  the radioactive decay  times, the timescales for the
radiative transport processes in the SN atmosphere, and the atomic
transitions would all have to vary as $(1+z)$ due to
variations in the masses of elementary particles for the theory
to account for the data. All these effects would have to result in making the observed spectra similar.

\section{Discussion}
\label{sect:discussion}

We have shown that a single stretch factor varying the timescale of
the SNe Ia accounts very well for the restframe $B$ band light curve
both before and after maximum light (up to 40 days past maximum).  It
is not understood from the current state of the theory of SNe Ia
whether this is a fortuitous coincidence or a reflection of some 
physics timescale
\citep{hoflich95,hoflich_khokhlov96,nuge_phd,pi2000,arnett00,hillebrandt}.
Although we do not necessarily expect this single timescale stretch factor to hold
for all wavelength bands, we do have evidence that it applies in the
$V$ band and (with less confidence) the $U$ band.   We will analyze
these results in future work.

The current analysis does not address the relationship between the SN
Ia light curve stretch and its peak luminosity.  This requires another
analysis that will be presented elsewhere.  However, it is important
to note that the current $B$ band template stretched by a factor $s$
fits the data as well as any parameterization can, given the current
data sets.  It will therefore yield a peak-luminosity correlation with
as small a dispersion as can be obtained by any other B-band
light-curve parameterization.

The comparison of the low-redshift and high-redshift composite
supernovae light curves in  Figure~\ref{template_fit} provide a first-order test for
evolution of supernovae as we go back to the $z \sim 0.5$ epoch.  The
close match of the light curves is suggestive that little evolution has
occurred.

A recent analysis of low-redshift early light curves by
Riess et al (1999a, 1999b) suggested that the explosion date of the
low-redshift SNe Ia was earlier than that of the high-redshift SNe.
The analyses of our early light curve uncertainty in Section 6 and in
AKN2000 show that the difference between the
two data sets is not very significant at this stage, less than
$\sim$2$\sigma$.  This will be an interesting region of the light curve
to pursue with future data, particularly over a range of host galaxy
environments that would be expected to show variations in metallicity
and progenitor ages. 
 We are currently working with the supernova research community
on generating a much more extensive low-redshift dataset to study this
question.
 However, it is important to remember that -- as pointed out by AKN2000 --
the template differences have a negligible influence on the corrected
peak magnitudes of the P99 SNe, and thus the cosmological parameters
derived therein are unchanged.

Finally, it is interesting to note that 
while the redshift of the light measures the expansion of the universe with a  ``microscopic clock" of period, typically $T =2\times10^{-15} $
seconds, our ``macroscopic clocks", the Type Ia SNe, measure the expansion over a $\approx 4$ week period or $T\approx 2.4\times10^6$ seconds. The $1 + z$ expansion effect is thus consistent for two time periods which differ by 21 orders of magnitude.

\vspace{0.5in}

\section{Acknowledgments}

The observations described in this paper
were primarily obtained as visiting/guest astronomers at
the Cerro Tololo
Inter-American Observatory 4-meter telescope, operated by the
National Optical Astronomy Observatory under contract to the National
Science Foundation; the Keck I and II 10-m telescopes of  the California
Association for Research in Astronomy;
the Wisconsin-Indiana-Yale-NOAO (WIYN) telescope;
the European Southern Observatory 3.6-meter telescope;
the Isaac
Newton and William Herschel Telescopes, operated by the Royal Greenwich
Observatory at the Spanish Observatorio del Roque de los Muchachos of
the Instituto de Astrofisica de Canarias;
the Hubble Space Telescope, and the Nordic Optical 2.5-meter
telescope.  We thank the dedicated staff of these observatories for
their excellent assistance in pursuit of this project.
In particular, Dianne Harmer, Paul
Smith and Daryl Willmarth were extraordinarily helpful as the WIYN queue
observers.   We thank Gary Bernstein and Tony Tyson for developing and
supporting the Big Throughput Camera at the CTIO 4-meter; this wide-field
camera was important in the discovery of many of the high-redshift supernovae.
This work was supported in part by the
Physics Division, E.~O. Lawrence Berkeley National Laboratory of the
U.~S. Department of Energy under Contract No. DE-AC03-76SF000098, and
by the National Science Foundation's Center for Particle Astrophysics,
University of California, Berkeley under grant No. ADT-88909616.
Support for this work was provided by NASA through grant number HST-GO-07336 from the Space Telescope Science Institute, which is operated by AURA, Inc., under NASA contract NAS5-26555.
A.~G. acknowledges the support of the Swedish Natural Science Research
Council.  The France-Berkeley Fund and the Stockholm-Berkeley Fund
provided additional collaboration support.



\vbox{
\footnotesize
\centerline{\vbox{\hsize 4.30in \noindent{\sc Table 1. Fit parameters
for 42 Type 1a Supernovae reported by the Supernova Cosmology Project}
(\cite{perl99}). 
For the present analysis we require good $R$-band photometry
and $K$ corrections to $B$, so the 7 objects with qualifying comments are
not used.}}
\vskip0.05in

\centerline{\vbox{
\halign{
        \hbox to 0.55in{\ \hss#\   } &   
        \hbox to 0.75in{\hss#\hss\ } &   
        \hbox to 0.50in{\hss#\hss\ } &   
        \hbox to 0.50in{\hss#\hss\ } &   
        \hbox to 0.50in{\hss#\hss\ } &   
        \hbox to 0.50in{\hss#\hss\ } &   
        \hbox to 0.50in{\hss#\hss\ } &   
        \hbox to 0.60in{\hss#      } \cr 
\noalign{\hrule\vskip2pt\hrule\smallskip}
Name & $z$ & $w$ & $\sigma_w$ & $s$ & $\sigma_s$ & $\chi^2/DoF$
 & Comment\rlap{s} \cr\noalign{\vskip-0.015in}
\noalign{\medskip\hrule\smallskip}
 1992bi  &  0.458  &   2.26  &   0.34  &   1.55  &   0.23  &   0.91  & \cr\noalign{\vskip-0.015in}
  1994F  &  0.354  &   0.96  &   0.19  &   0.71  &   0.14  &   1.74  & \cr\noalign{\vskip-0.015in}
  1994G  &  0.425  &   1.32  &   0.24  &   0.92  &   0.17  &   0.71  & Late $R^*$ \cr\noalign{\vskip-0.015in}
 1994H   &  0.374  &   1.19  &   0.07  &   0.87  &   0.05  &   1.59  & \cr\noalign{\vskip-0.015in}
 1994al  &  0.420  &   1.22  &   0.13  &   0.86  &   0.09  &   0.94  & \cr\noalign{\vskip-0.015in}
 1994am  &  0.372  &   1.22  &   0.05  &   0.89  &   0.04  &   1.73  & \cr\noalign{\vskip-0.015in}
 1994an  &  0.378  &   1.44  &   0.23  &   1.04  &   0.17  &   1.38  & \cr\noalign{\vskip-0.015in}
 1995aq  &  0.453  &   1.27  &   0.15  &   0.87  &   0.10  &   1.21  & \cr\noalign{\vskip-0.015in}
 1995ar  &  0.465  &   1.42  &   0.21  &   0.97  &   0.14  &   2.06  & \cr\noalign{\vskip-0.015in}
 1995as  &  0.498  &   1.64  &   0.16  &   1.09  &   0.11  &   0.57  & \cr\noalign{\vskip-0.015in}
 1995at  &  0.655  &   1.84  &   0.12  &   1.11  &   0.07  &   0.80  & \cr\noalign{\vskip-0.015in}
 1995aw  &  0.400  &   1.62  &   0.06  &   1.16  &   0.04  &   1.80  & \cr\noalign{\vskip-0.015in}
 1995ax  &  0.615  &   1.88  &   0.18  &   1.16  &   0.11  &   1.17  & \cr\noalign{\vskip-0.015in}
 1995ay  &  0.480  &   1.36  &   0.12  &   0.92  &   0.08  &   1.07  & \cr\noalign{\vskip-0.015in}
 1995az  &  0.450  &   1.41  &   0.10  &   0.97  &   0.07  &   0.63  & \cr\noalign{\vskip-0.015in}
 1995ba  &  0.388  &   1.36  &   0.06  &   0.98  &   0.04  &   1.13  & \cr\noalign{\vskip-0.015in}
 1996cf  &  0.570  &   1.61  &   0.11  &   1.03  &   0.07  &   1.06  & \cr\noalign{\vskip-0.015in}
 1996cg  &  0.490  &   1.58  &   0.07  &   1.06  &   0.05  &   0.85  & \cr\noalign{\vskip-0.015in}
 1996ci  &  0.495  &   1.53  &   0.07  &   1.02  &   0.05  &   1.21  & \cr\noalign{\vskip-0.015in}
 1996ck  &  0.656  &   1.51  &   0.20  &   0.91  &   0.12  &   1.06  & \cr\noalign{\vskip-0.015in}
 1996cl  &  0.828  &   2.07  &   0.53  &   1.13  &   0.29  &   1.12  & $U$ band \cr\noalign{\vskip-0.015in}
 1996cm  &  0.450  &   1.33  &   0.09  &   0.92  &   0.06  &   0.80  & \cr\noalign{\vskip-0.015in}
 1996cn  &  0.430  &   1.28  &   0.10  &   0.89  &   0.07  &   1.30  & \cr\noalign{\vskip-0.015in}
  1997F  &  0.580  &   1.62  &   0.11  &   1.02  &   0.07  &   0.86  & \cr\noalign{\vskip-0.015in}
  1997G  &  0.763  &   1.71  &   0.30  &   0.97  &   0.17  &   1.17  & $U$ band \cr\noalign{\vskip-0.015in}  
  1997H  &  0.526  &   1.38  &   0.08  &   0.90  &   0.05  &   0.86  & \cr\noalign{\vskip-0.015in}
  1997I  &  0.172  &   1.11  &   0.04  &   0.94  &   0.03  &   5.60$^{**}$   & $V$ band \cr\noalign{\vskip-0.015in}
  1997J  &  0.619  &   1.63  &   0.21  &   1.00  &   0.13  &   0.94  & \cr\noalign{\vskip-0.015in}
  1997K  &  0.592  &   1.87  &   0.30  &   1.18  &   0.19  &   0.62  & \cr\noalign{\vskip-0.015in}
  1997L  &  0.550  &   1.51  &   0.14  &   0.98  &   0.09  &   2.57  & \cr\noalign{\vskip-0.015in}
  1997N  &  0.180  &   1.21  &   0.02  &   1.03  &   0.02  &   1.72  & $V$ band \cr\noalign{\vskip-0.015in}
  1997O  &  0.374  &   1.40  &   0.10  &   1.02  &   0.07  &   0.75  &  Outlier \cr\noalign{\vskip-0.015in}
  1997P  &  0.472  &   1.40  &   0.06  &   0.95  &   0.04  &   1.30  & \cr\noalign{\vskip-0.015in}
  1997Q  &  0.430  &   1.36  &   0.04  &   0.95  &   0.03  &   1.09  & \cr\noalign{\vskip-0.015in}
  1997R  &  0.657  &   1.65  &   0.12  &   0.99  &   0.07  &   1.06  & \cr\noalign{\vskip-0.015in}
  1997S  &  0.612  &   1.90  &   0.10  &   1.18  &   0.06  &   1.66  & \cr\noalign{\vskip-0.015in}
 1997ac  &  0.320  &   1.39  &   0.03  &   1.05  &   0.02  &   0.79  & \cr\noalign{\vskip-0.015in}
 1997af  &  0.579  &   1.39  &   0.08  &   0.88  &   0.05  &   0.95  & \cr\noalign{\vskip-0.015in}
 1997ai  &  0.450  &   1.52  &   0.20  &   1.04  &   0.14  &   0.82  & \cr\noalign{\vskip-0.015in}
 1997aj  &  0.581  &   1.49  &   0.09  &   0.94  &   0.06  &   1.75  & \cr\noalign{\vskip-0.015in}
 1997am  &  0.416  &   1.55  &   0.07  &   1.10  &   0.05  &   1.24  & \cr\noalign{\vskip-0.015in}
 1997ap  &  0.830  &   1.88  &   0.09  &   1.03  &   0.05  &   1.04  & $U$ band \cr\noalign{\vskip-0.015in}
\noalign{\vskip0.05in}
\noalign{\hrule\vskip2pt\hrule\smallskip}\cr\noalign{\vskip-0.015in}
\multispan7 * No $R$-band data before 14 days; primary photometry from $I$ band. & \cr
\multispan8 ** Here the V template was used while a template between V and R would be more \cr
\multispan6    \hskip0.2in appropriate. A fit to Parab-18 gives $\chi^2/DoF = 2.8$. & &\cr
}}}}

\clearpage
\footnotesize
\centerline{\sc Table 2. Fit parameters for the Cal\'{a}n/Tololo data}
\vskip0.05in
\centerline{\vbox{
\halign{
        \hbox to 0.55in{\ \hss#\   } &  
        \hbox to 0.75in{\hss#\hss\ } &  
        \hbox to 0.50in{\hss#\hss\ } &  
        \hbox to 0.50in{\hss#\hss\ } &  
        \hbox to 0.50in{\hss#\hss\ } &  
        \hbox to 0.50in{\hss#\hss\ } &  
        \hbox to 0.50in{\hss#} \cr  
\noalign{\hrule\vskip2pt\hrule\smallskip}
Name & $z$ & $w$ & $\sigma_w$ & $s$ & $\sigma_s$  &  $\chi^2/DoF $\cr\noalign{\vskip-0.015in}
\noalign{\medskip\hrule\smallskip}
   1990O  &  0.030  &   1.09  &   0.03  &   1.06  &   0.03  &   1.53  \cr\noalign{\vskip-0.015in}
  1990af  &  0.050  &   0.82  &   0.02  &   0.78  &   0.02  &   0.51  \cr\noalign{\vskip-0.015in}
   1992P  &  0.026  &   1.15  &   0.08  &   1.12  &   0.08  &   1.45  \cr\noalign{\vskip-0.015in}
  1992ae  &  0.075  &   1.09  &   0.09  &   1.02  &   0.08  &   0.78  \cr\noalign{\vskip-0.015in}
  1992ag  &  0.026  &   1.14  &   0.04  &   1.11  &   0.04  &   2.15  \cr\noalign{\vskip-0.015in}
  1992al  &  0.014  &   0.99  &   0.02  &   0.98  &   0.02  &   1.71  \cr\noalign{\vskip-0.015in}
  1992aq  &  0.101  &   1.04  &   0.14  &   0.95  &   0.13  &   0.65  \cr\noalign{\vskip-0.015in}
  1992bc  &  0.020  &   1.12  &   0.01  &   1.09  &   0.01  &   3.14  \cr\noalign{\vskip-0.015in}
  1992bg  &  0.036  &   1.09  &   0.05  &   1.05  &   0.05  &   1.68  \cr\noalign{\vskip-0.015in}
  1992bh  &  0.045  &   1.15  &   0.05  &   1.10  &   0.05  &   3.77  \cr\noalign{\vskip-0.015in}
  1992bl  &  0.043  &   0.92  &   0.03  &   0.88  &   0.03  &   2.15  \cr\noalign{\vskip-0.015in}
  1992bo  &  0.018  &   0.77  &   0.01  &   0.76  &   0.01  &   1.29  \cr\noalign{\vskip-0.015in}
  1992bp  &  0.079  &   1.03  &   0.03  &   0.95  &   0.03  &   1.35  \cr\noalign{\vskip-0.015in}
  1992br  &  0.088  &   0.58  &   0.04  &   0.53  &   0.04  &   1.39  \cr\noalign{\vskip-0.015in}
  1992bs  &  0.063  &   1.05  &   0.05  &   0.99  &   0.05  &   1.43  \cr\noalign{\vskip-0.015in}
   1993B  &  0.071  &   1.06  &   0.09  &   0.99  &   0.08  &   1.14  \cr\noalign{\vskip-0.015in}
   1993O  &  0.052  &   0.99  &   0.01  &   0.94  &   0.01  &   1.26  \cr\noalign{\vskip-0.015in}
  1993ag  &  0.050  &   1.01  &   0.04  &   0.96  &   0.04  &   1.16  \cr\noalign{\vskip-0.015in}
\noalign{\vskip0.05in}
\noalign{\hrule\vskip2pt\hrule\smallskip}
}}}

\def\z{\phantom0}
\footnotesize
\vbox{
\centerline{\sc Table 3. $B$ templates, in Normalized Flux, used by the Supernova Cosmology Project}
\noindent
The headers: ``SCP1997'' refers to \cite{leibundgutthesis} from day -5 on up, with a linear extension (in magnitude) to earlier times shown in flux here. As discussed in the text, this is the template used in  \cite{perl97,perl99}. Tables 1, 2 and 7 as well as Figure 2(e), are based on this template. ``Parab - 18" uses a parabolic turn-on at $-17.6$~days from a fit to the 35 SCP SNe. Tables 4 and 6 and Figures 1 and 2, except 2(c) and 2(e), are based on this template. ``Parab -20" uses the parabolic turn-on at -20 days following the early rise-time data of \cite{riess_risea}. Table 5 and  Figure 2(c) are based on this template. After -10 days the two parabolic templates coincide.
\vskip0.08in
      \vbox{\hbox to1.0\hsize{%
          \vtop{\hsize=0.48\hsize\noindent
\vbox{
\halign{
\hbox to 0.55in{\ \hss#\ }&
        \hbox to 0.75in{\hss#\hss\ } &
        \hbox to 0.75in{\hss#\hss\ } &
        \hbox to 0.75in{\hss#\hss\ } \cr
\noalign{\hrule\vskip2pt\hrule\smallskip}
& ~SCP1997 & ~~Parab~~ & ~~Parab~~ \cr\noalign{\vskip-0.015in}
Day & ~~~~~~~~~~ &  -18~~ & -20 \cr
\noalign{\smallskip\hrule\smallskip}
  -25 &    0.001 &    0.000 &    0.000 \cr\noalign{\vskip-0.015in}
  -24 &    0.001 &    0.000 &    0.000 \cr\noalign{\vskip-0.015in}
  -23 &    0.002 &    0.000 &    0.000 \cr\noalign{\vskip-0.015in}
  -22 &    0.003 &    0.000 &    0.000 \cr\noalign{\vskip-0.015in}
  -21 &    0.005 &    0.000 &    0.000 \cr\noalign{\vskip-0.015in}
  -20 &    0.007 &    0.000 &    0.000 \cr\noalign{\vskip-0.015in}
  -19 &    0.012 &    0.000 &    0.005 \cr\noalign{\vskip-0.015in}
  -18 &    0.019 &    0.000 &    0.019 \cr\noalign{\vskip-0.015in}
  -17 &    0.030 &    0.005 &    0.044 \cr\noalign{\vskip-0.015in}
  -16 &    0.048 &    0.025 &    0.080 \cr\noalign{\vskip-0.015in}
  -15 &    0.076 &    0.063 &    0.127 \cr\noalign{\vskip-0.015in}
  -14 &    0.120 &    0.118 &    0.183 \cr\noalign{\vskip-0.015in}
  -13 &    0.189 &    0.190 &    0.248 \cr\noalign{\vskip-0.015in}
  -12 &    0.277 &    0.279 &    0.324 \cr\noalign{\vskip-0.015in}
  -11 &    0.388 &    0.385 &    0.410 \cr\noalign{\vskip-0.015in}
  -10 &    0.494 &    0.508 &    0.508 \cr\noalign{\vskip-0.015in}
\noalign{\vskip0.05in\hrule\vskip0.05in}
   -9 &    0.597 &    \multispan2{  0.622} \cr\noalign{\vskip-0.015in}
   -8 &    0.674 &    \multispan2{  0.714} \cr\noalign{\vskip-0.014in}
   -7 &    0.746 &    \multispan2{  0.789} \cr\noalign{\vskip-0.015in}
   -6 &    0.810 &    \multispan2{  0.849} \cr\noalign{\vskip-0.015in}
   -5 &    0.868 &    \multispan2{  0.896} \cr\noalign{\vskip-0.015in}
   -4 &    0.915 &    \multispan2{  0.933} \cr\noalign{\vskip-0.015in}
   -3 &    0.952 &    \multispan2{  0.961} \cr\noalign{\vskip-0.015in}
   -2 &    0.979 &    \multispan2{  0.982} \cr\noalign{\vskip-0.015in}
   -1 &    0.995 &    \multispan2{  0.995} \cr\noalign{\vskip-0.015in}
    0 &    1.000 &    \multispan2{  1.000} \cr\noalign{\vskip-0.015in}
    1 &    0.995 &    \multispan2{  0.995} \cr\noalign{\vskip-0.015in}
    2 &    0.979 &    \multispan2{  0.978} \cr\noalign{\vskip-0.015in}
    3 &    0.953 &    \multispan2{  0.950} \cr\noalign{\vskip-0.015in}
    4 &    0.918 &    \multispan2{  0.910} \cr\noalign{\vskip-0.015in}
    5 &    0.875 &    \multispan2{  0.863} \cr\noalign{\vskip-0.015in}
    6 &    0.827 &    \multispan2{  0.811} \cr\noalign{\vskip-0.015in}
    7 &    0.773 &    \multispan2{  0.759} \cr\noalign{\vskip-0.015in}
    8 &    0.718 &    \multispan2{  0.707} \cr\noalign{\vskip-0.015in}
    9 &    0.661 &    \multispan2{  0.656} \cr\noalign{\vskip-0.015in}
   10 &    0.604 &    \multispan2{  0.606} \cr\noalign{\vskip-0.015in}
   11 &    0.550 &    \multispan2{  0.558} \cr\noalign{\vskip-0.015in}
   12 &    0.498 &    \multispan2{  0.511} \cr\noalign{\vskip-0.015in}
\noalign{\vskip0.05in}
\noalign{\hrule\vskip2pt\hrule\smallskip}
}}
\hfil}\hfill%
          \vtop{\hsize=0.48\hsize\noindent
\vbox{
\halign{ 
\hbox to 0.55in{\ \phantom{$-$}#\hss \ }&
        \hbox to 0.75in{\hss#\hss\ } &
        \hbox to 0.75in{\hss#\hss\ } \cr 
\noalign{\hrule\vskip2pt\hrule\smallskip}
& ~SCP1997 & ~~Parab~~ \cr\noalign{\vskip-0.015in} 
Day & ~~~~~~~~~ & -18~~ \& -20 \cr 
\noalign{\smallskip\hrule\smallskip}
   13 &    0.449 &    0.466 \cr\noalign{\vskip-0.015in}
   14 &    0.404 &    0.423 \cr\noalign{\vskip-0.015in}
   15 &    0.363 &    0.381 \cr\noalign{\vskip-0.015in}
   16 &    0.327 &    0.342 \cr\noalign{\vskip-0.015in}
   17 &    0.293 &    0.306 \cr\noalign{\vskip-0.015in}
   18 &    0.264 &    0.272 \cr\noalign{\vskip-0.015in}
   19 &    0.238 &    0.241 \cr\noalign{\vskip-0.015in}
   20 &    0.215 &    0.215 \cr\noalign{\vskip-0.015in}
   21 &    0.195 &    0.192 \cr\noalign{\vskip-0.015in}
   22 &    0.178 &    0.172 \cr\noalign{\vskip-0.015in}
   23 &    0.162 &    0.155 \cr\noalign{\vskip-0.015in}
   24 &    0.149 &    0.142 \cr\noalign{\vskip-0.015in}
   25 &    0.137 &    0.130 \cr\noalign{\vskip-0.015in}
   26 &    0.127 &    0.120 \cr\noalign{\vskip-0.015in}
   27 &    0.118 &    0.112 \cr\noalign{\vskip-0.015in}
   28 &    0.110 &    0.105 \cr\noalign{\vskip-0.015in}
   29 &    0.103 &    0.099 \cr\noalign{\vskip-0.015in}
   30 &    0.097 &    0.094 \cr\noalign{\vskip-0.015in}
   31 &    0.092 &    0.090 \cr\noalign{\vskip-0.015in}
   32 &    0.087 &    0.087 \cr\noalign{\vskip-0.015in}
   33 &    0.083 &    0.084 \cr\noalign{\vskip-0.015in}
   34 &    0.080 &    0.081 \cr\noalign{\vskip-0.015in}
   35 &    0.076 &    0.078 \cr\noalign{\vskip-0.015in}
   36 &    0.074 &    0.076 \cr\noalign{\vskip-0.015in}
   37 &    0.071 &    0.073 \cr\noalign{\vskip-0.015in}
   38 &    0.069 &    0.071 \cr\noalign{\vskip-0.015in}
   39 &    0.067 &    0.069 \cr\noalign{\vskip-0.015in}
   40 &    0.065 &    0.067 \cr\noalign{\vskip-0.015in}
   41 &    0.064 &    0.065 \cr\noalign{\vskip-0.015in}
   42 &    0.062 &    0.064 \cr\noalign{\vskip-0.015in}
   43 &    0.061 &    0.062 \cr\noalign{\vskip-0.015in}
   44 &    0.060 &    0.060 \cr\noalign{\vskip-0.015in}
   45 &    0.059 &    0.059 \cr\noalign{\vskip-0.015in}
   46 &    0.058 &    0.058 \cr\noalign{\vskip-0.015in}
   47 &    0.057 &    0.057 \cr\noalign{\vskip-0.015in}
   48 &    0.056 &    0.056 \cr\noalign{\vskip-0.015in}
   49 &    0.055 &    0.055 \cr\noalign{\vskip-0.015in}
   50 &    0.054 &    0.054 \cr\noalign{\vskip-0.015in}
\noalign{\vskip0.05in}
\noalign{\hrule\vskip2pt\hrule\smallskip}
}}
\hfil}
}}%
}

\clearpage
\footnotesize
\centerline{\sc Table 4. $\chi^2$ Fits of the SCP Data to the}
\centerline{\sc Composite Parabolic Template  ``Parab-18''}
\centerline{\sc  Corresponding to $t_{exp} = -17.6$ Days}
\vskip0.02in
\centerline{Columns 1 and 2 define the Epoch Interval,}
\centerline{Columns 3 and 4 give the $\chi^2$ and }
\centerline{$\chi^2$ per Degree of Freedom. See Fig 2 (b)}
\vskip0.05in
\centerline{\vbox{
\halign{
        \hbox to 0.5in{\ \hss#\             } &   
        \hbox to 0.5in{\hss#\hss\hskip0.10in} &   
        \hbox to 0.5in{\hss#\hss\           } &   
        \hbox to 0.5in{\hss#                } \cr 
\noalign{\hrule\vskip2pt\hrule\smallskip}
\multispan2{\hss Epoch\hss} &   $\chi^2$ & \omit{\hss $\chi^2/$DoF}  \cr
\noalign{\vskip-0.015in}
\omit{\hss Start\hskip0.01in} & \omit{\hskip0.1in End\hskip0.01in} \cr
\noalign{\vskip-0.015in}
\noalign{\medskip\hrule\smallskip}  
   -25  & -20  &   27.8   &  1.16  \cr\noalign{\vskip-0.015in}
   -20  & -15  &   23.7   &  0.99  \cr\noalign{\vskip-0.015in}
   -15  & -10  &   35.8   &  0.58  \cr\noalign{\vskip-0.015in}
   -10  &  -5  &   72.4   &  0.95  \cr\noalign{\vskip-0.015in}
    -5  &   0  &  101.7   &  1.02  \cr\noalign{\vskip-0.015in}
     0  &   5  &  111.6   &  0.88  \cr\noalign{\vskip-0.015in}
     5  &  10  &   74.5   &  0.89  \cr\noalign{\vskip-0.015in}
    10  &  15  &   62.2   &  0.91  \cr\noalign{\vskip-0.015in}
    15  &  20  &   25.6   &  0.47  \cr\noalign{\vskip-0.015in}
    20  &  25  &   83.3   &  0.95  \cr\noalign{\vskip-0.015in}
    25  &  30  &   70.1   &  1.02  \cr\noalign{\vskip-0.015in}
    30  &  35  &   49.1   &  0.88  \cr\noalign{\vskip-0.015in}
    35  &  40  &   25.8   &  1.51  \cr\noalign{\vskip-0.015in}
\noalign{\medskip\hrule\smallskip} 
   -25  &  40  &  763.5   &  0.905 \cr\noalign{\vskip-0.015in}
\noalign{\vskip0.05in}
\noalign{\hrule\vskip2pt\hrule\smallskip}
}}}
\vskip0.2in
\footnotesize
\centerline{\sc Table 5. $\chi^2$ Fits of the SCP Data to the}
\centerline{\sc Composite Parabolic Template  ``Parab-20''}
\centerline{\sc  Corresponding to $t_{exp} = -20$ Days}
\vskip0.02in
\centerline{Column headings as in Table 4. See Figure 2(d) }
\vskip0.05in
\centerline{\vbox{
\halign{
        \hbox to 0.5in{\ \hss#\   } &   
        \hbox to 0.5in{\hss#\hss\hskip0.10in}&  
        \hbox to 0.5in{\hss#\hss\ } &  
        \hbox to 0.5in{\hss#      } \cr  
\noalign{\hrule\vskip2pt\hrule\smallskip}
\multispan2{\hss Epoch\hss} &   $\chi^2$ & \omit{\hss $\chi^2/$DoF}  \cr
\noalign{\vskip-0.015in}
\omit{\hss Start\hskip0.01in} & \omit{\hskip0.1in End\hskip0.01in} & & \cr
\noalign{\vskip-0.015in}
\noalign{\medskip\hrule\smallskip}  
   -25  & -20  &   29.3   &  1.01   \cr\noalign{\vskip-0.015in}
   -20  & -15  &   27.5   &  0.65   \cr\noalign{\vskip-0.015in}
   -15  & -10  &   28.4   &  0.71   \cr\noalign{\vskip-0.015in}
   -10  &  -5  &   71.2   &  0.94   \cr\noalign{\vskip-0.015in}
    -5  &   0  &  105.0   &  1.00   \cr\noalign{\vskip-0.015in}
     0  &   5  &  115.6   &  0.92   \cr\noalign{\vskip-0.015in}
     5  &  10  &   70.1   &  0.86   \cr\noalign{\vskip-0.015in}
    10  &  15  &   63.6   &  0.96   \cr\noalign{\vskip-0.015in}
    15  &  20  &   26.1   &  0.49   \cr\noalign{\vskip-0.015in}
    20  &  25  &   84.6   &  0.95   \cr\noalign{\vskip-0.015in}
    25  &  30  &   67.9   &  1.03   \cr\noalign{\vskip-0.015in}
    30  &  35  &   52.0   &  0.91   \cr\noalign{\vskip-0.015in}
    35  &  40  &   21.4   &  1.34   \cr\noalign{\vskip-0.015in}
\noalign{\medskip\hrule\smallskip} 
   -25  &  40  &  762.6   &  0.906   \cr\noalign{\vskip-0.015in}
\noalign{\vskip0.05in}
\noalign{\hrule\vskip2pt\hrule\smallskip}
}}}

\clearpage
 
\vskip0.5in
\footnotesize
\centerline{\sc Table 6. $\chi^2$ Fits of the Cal\'{a}n/Tololo data}
\centerline{\sc to the Composite Parabolic Template$^*$ }
\vskip0.02in
\centerline{Column headings as in Table 4. See Figure 2(c) }
\vskip0.05in
\centerline{\vbox{
\halign{
        \hbox to 0.5in{\ \hss#\hskip0.10in   } &             
        \hbox to 0.5in{\hss#\hss\hskip0.10in } &  
        \hbox to 0.5in{\hss#\hss\            } &  
        \hbox to 0.5in{\hss#\hss\            } \cr  
\noalign{\hrule\vskip2pt\hrule\smallskip}
\multispan2{\hss Epoch\hss} &   $\chi^2$ & \omit{\hskip0.03in $\chi^2/$DoF} \cr
\omit{\hss Start\hskip0.01in} & \omit{\hskip0.1in End\hskip0.01in} & \cr
\noalign{\vskip-0.015in}
\noalign{\medskip\hrule\smallskip}
   -10  &  -5  &   25.3   &  1.95   \cr\noalign{\vskip-0.015in}
    -5  &   0  &   39.7   &  2.09    \cr\noalign{\vskip-0.015in}
     0  &   5  &   17.1   &  0.47     \cr\noalign{\vskip-0.015in}
     5  &  10  &   59.5   &  1.06    \cr\noalign{\vskip-0.015in}
    10  &  15  &   55.7   &  1.39     \cr\noalign{\vskip-0.015in}
    15  &  20  &   28.9   &  1.25    \cr\noalign{\vskip-0.015in}
    20  &  25  &   14.8   &  0.93   \cr\noalign{\vskip-0.015in}
    25  &  30  &   11.2   &  0.62    \cr\noalign{\vskip-0.015in}
    30  &  35  &   21.1   &  1.32   \cr\noalign{\vskip-0.015in}
    35  &  40  &   17.0   &  0.81    \cr\noalign{\vskip-0.015in}
 \noalign{\medskip\hrule\smallskip} 
   -10  &  40  &  290.3  &  1.143  \cr\noalign{\vskip-0.015in}
 \noalign{\vskip0.05in}
\noalign{\hrule\vskip2pt\hrule\smallskip}
\multispan4 * In this Table some of the photometry \cr 
\multispan4 \hskip0.2in errors were extremely small.  We used \cr 
\multispan4 \hskip0.2in a minimum error of 0.007, in \hskip0.1in \cr
\multispan3 \hskip0.2in normalized units, here. & \cr
}}}
\vskip0.1in
\footnotesize
\vbox{
\centerline{\sc Table 7. $\chi^2$ Fits of the SCP Data to the}
\centerline{\sc Composite Exponential Template  ``SCP1997''}
\centerline{\sc  the Extended Leibundgut Template}
\vskip0.02in
\centerline{Column headings as in Table 4. See Figure 2(e) }
\vskip0.05in
\centerline{\vbox{
\halign{
        \hbox to 0.5in{\ \hss#\   } &   
        \hbox to 0.5in{\hss#\hss\hskip0.10in}&  
        \hbox to 0.5in{\hss#\hss\ } &  
        \hbox to 0.5in{\hss#\hss\ } \cr  
\noalign{\hrule\vskip2pt\hrule\smallskip}
\multispan2{\hss Epoch\hss} &   $\chi^2$ & \omit{\hss $\chi^2/$DoF}  \cr
\noalign{\vskip-0.015in}
\omit{\hss Start\hskip0.01in} & \omit{\hskip0.1in End\hskip0.01in} \cr
\noalign{\vskip-0.015in}
\noalign{\medskip\hrule\smallskip}  
   -25  & -20  &   27.5   &  1.15    \cr\noalign{\vskip-0.015in}
   -20  & -15  &   24.3   &  1.01    \cr\noalign{\vskip-0.015in}
   -15  & -10  &   36.8   &  0.59   \cr\noalign{\vskip-0.015in}
   -10  &  -5  &  105.2   &  1.38   \cr\noalign{\vskip-0.015in}
    -5  &   0  &  105.7   &  1.06   \cr\noalign{\vskip-0.015in}
     0  &   5  &  110.8   &  0.87     \cr\noalign{\vskip-0.015in}
     5  &  10  &   75.3   &  0.90     \cr\noalign{\vskip-0.015in}
    10  &  15  &   73.4   &  1.08     \cr\noalign{\vskip-0.015in}
    15  &  20  &   29.3   &  0.54    \cr\noalign{\vskip-0.015in}
    20  &  25  &   83.8   &  0.95    \cr\noalign{\vskip-0.015in}
    25  &  30  &   69.1   &  1.00    \cr\noalign{\vskip-0.015in}
    30  &  35  &   49.2   &  0.88     \cr\noalign{\vskip-0.015in}
    35  &  40  &   25.4   &  1.49    \cr\noalign{\vskip-0.015in}
\noalign{\medskip\hrule\smallskip} 
   -25  &  40  &  815.7  &  0.966    \cr\noalign{\vskip-0.015in}
\noalign{\vskip0.05in}
\noalign{\hrule\vskip2pt\hrule\smallskip}
}}}
}

\clearpage

\vskip0.5in
\footnotesize
\centerline{\sc Table 8. $\chi^2$ Fits of the SCP Data to the}
\centerline{\sc Parabolic Template ``Parab-18''}
\centerline{\sc with only $1 + z$ but not $s$ factored out}
\vskip0.02in
\centerline{This Table illustrates the effect of the  }
\centerline{factor $s$ both before and after Maximum light.}
\centerline{Column headings as in Table 4.}
\vskip0.05in
\centerline{\vbox{
\halign{
        \hbox to 0.5in{\ \hss#\ } &   
        \hbox to 0.5in{\hss#\hss\hskip0.10in}&  
        \hbox to 0.5in{\hss#\hss\ } &  
        \hbox to 0.5in{\hss#\hss\ } \cr  
\noalign{\hrule\vskip2pt\hrule\smallskip}
\multispan2{\hss Epoch\hss}   & $\chi^2$ & \omit{\hss $\chi^2/$DoF}   \cr
\noalign{\vskip-0.015in}
\omit{\hss Start\hskip0.01in} & \omit{\hskip0.1in End\hskip0.01in} \cr
\noalign{\vskip-0.015in}
\noalign{\medskip\hrule\smallskip}  
   -25  & -20  &   24.2   &  1.86  \cr\noalign{\vskip-0.015in}
   -20  & -15  &   30.3   &  1.08  \cr\noalign{\vskip-0.015in}
   -15  & -10  &  231.8   &  4.55  \cr\noalign{\vskip-0.015in}
   -10  &  -5  &  250.2   &  3.05  \cr\noalign{\vskip-0.015in}
    -5  &   0  &  104.3   &  1.03  \cr\noalign{\vskip-0.015in}
     0  &   5  &  112.4   &  0.89  \cr\noalign{\vskip-0.015in}
     5  &  10  &  125.6   &  1.38  \cr\noalign{\vskip-0.015in}
    10  &  15  &   57.9   &  0.95  \cr\noalign{\vskip-0.015in}
    15  &  20  &  110.1   &  1.84  \cr\noalign{\vskip-0.015in}
    20  &  25  &   98.6   &  1.15  \cr\noalign{\vskip-0.015in}
    25  &  30  &  105.0   &  1.36  \cr\noalign{\vskip-0.015in}
    30  &  35  &   45.9   &  1.12  \cr\noalign{\vskip-0.015in}
    35  &  40  &   35.4   &  1.18  \cr\noalign{\vskip-0.015in}
 \noalign{\medskip\hrule\smallskip} 
   -25  &  40  & 1330.7  &  1.582   \cr\noalign{\vskip-0.015in}
\noalign{\vskip0.05in}
\noalign{\hrule\vskip2pt\hrule\smallskip}
}}}


\bibliography{refs}

\clearpage
\begin{thebibliography}{}



\bibitem[Aldering , Knop \& Nugent(1999)[Aldering , Knop \& Nugent]{AKN99}
Aldering, G., Knop R. \& Nugent P. 2000, AJ, 119, 2110 




\bibitem[Arnett(2000)Arnett]{arnett00}
Arnett, W.~D. 1999, astro-ph/990903

\bibitem[Arp(1987)Arp]{arp87}
Arp, H.C. 1987, Quasars, Redshifts and Controversies (Interstellar
Media, Berkeley)

\bibitem[Arp et al.(1990)Arp et al.]{arp90}
Arp, H.C., Burbidge, G., Hoyle, F., Narkikar, J. V., \& 
Wickramasinghe, N.~C. 1990, Nature, 346, 807

\bibitem[Arp et al.(1994)Arp et al.]{arp94}
Arp, H.C., et al. 1994, ApJ, 430, 74




\bibitem[Chow(1977)Chow]{chow81}
Chow, T.~L. 1981, Lett.\ Nuovo Cimento, 32, 351



 

\bibitem[Goldhaber {et~al.}(1995)Goldhaber {\em et~al.}]{goldhaberaigua}
Goldhaber, G., et~al. 1995,
in Presentations at the NATO ASI in Aiguablava, Spain, LBL-38400,
  page III.1; also published in Thermonuclear Supernova, P.~Ruiz-Lapuente,
  R.~Canal, and J.Isern, editors, Dordrecht: Kluwer, page 777 (1997)

\bibitem[Goldhaber(1998a)Goldhaber]{goldhaberslac}
Goldhaber, G. 1998a, XXVI SLAC Summer Institute "Gravity: From the Hubble Length to the Plank Length" Aug 1998

\bibitem[Goldhaber(1998b)Goldhaber]{goldhaberaas}
Goldhaber, G. 1998b, B.A.A.S., 193, 47.13
 
\bibitem[Groom(1998)Groom]{groomaas}
Groom, D.~E. 1998, B.A.A.S., 193, 111.02


\bibitem[Hamuy {et~al.}(1995)Hamuy, Phillips, Maza, Suntzeff, Schommer, and
  Aviles]{hametal95}
Hamuy, M., Phillips, M.~M., Maza, J., Suntzeff, N.~B., Schommer, R.~A., \&
  Aviles, R. 1995, AJ, 109, 1

\bibitem[Hamuy {et~al.}(1996)Hamuy, Phillips, Maza, Suntzeff, Schommer, and
  Aviles]{hametal96}
Hamuy, M., Phillips, M.~M., Maza, J., Suntzeff, N.~B., Schommer, R.~A., \&
  Aviles, R. 1996, AJ, 112, 2391

\bibitem[Hillebrandt \& Niemeyer(2000)Hillebrandt and Niemeyer]{hillebrandt}
Hillebrandt W., \& Niemeyer J.~C., 2000, A R A \& A in press (astro-ph/0006305)

\bibitem[H\"oflich(1995)H\"oflich]{hoflich95}
H\"oflich, P. 1995, \apj, 443, 89

\bibitem[H\"oflich \& Khokhlov(1996)H\"oflich and Khokhlov]{hoflich_khokhlov96}
H\"oflich, P.,  \& Khokhlov, A.~M. 1996, \apj, 457, 500


\bibitem[Hoyle \& Fowler(1960)Hoyle \& Fowler]{hoyle60}
Hoyle, F., \& Fowler, W.~A. 1960, \apj, 132, 565

\bibitem[Hubble \& Tolman(1935)Hubble \& Tolman]{hubble_tolman35} 
Hubble, E., and Tolman, R.~C. 1935, \apj, 82, 302

\bibitem[Hubble(1936)Hubble]{hubble36} 
Hubble, E. 1936, \apj, 84, 517

\bibitem[James \& Roos(1994)James \& Roos]{MINUIT}
James, F., \&  Roos, M.,  ``MINUIT, Function Minimization and Error
Analysis," CERN D506 (Long Writeup). Available from the CERN Program
Library Office, CERN-IT Division, CERN, CH-1211, Geneva 21, Switzerland


\bibitem[Kim, Goobar, \& Perlmutter(1996)Kim, Goobar, and
Perlmutter]{kim_kcorr96} Kim, A., Goobar, A., \& Perlmutter, S. 1996,
PASP, 108,  190



\bibitem[Lee {et~al.}(2000)Lee{\em et~al.}]{lee2000}
Lee,~A., Bloom~E.~D. \&  Petrosian~V. 2000, astro-ph/0002218


\bibitem[Leibundgut(1988)Leibundgut]{leibundgutthesis}
Leibundgut, B., 1988, Ph.~D. Thesis, Univesity of Basel

\bibitem[Leibundgut {et~al.}(1991)Leibundgut {\em et~al.}]{leib1990N}
Leibundgut, B., et~al. 1991, \apj, 371, L23

\bibitem[Leibundgut {et~al.}(1991)Leibundgut {\em et~al.}]{leibundgut91}
Leibundgut, B., Tammann, G.~A., Cadonau, R., \& Cerrito, D. 1991,
Astron.\ Astrophys.\ Suppl.\ Ser., 89, 537

\bibitem[Leibundgut {et~al.}(1996)Leibundgut {\em et~al.}]{leibundgutdilation}
Leibundgut, B., et~al. 1996, \apj, 466, L21


\bibitem[Lentz {et al}.(2000)Lentz {\em et al.}]{lentz99}
Lentz, E.~ J., Baron, E., Branch, D. , Hauschildt, P.~ H., \& Nugent, P.~ E.
2000, ApJ., 530, 966

\bibitem[Lira {et al}.(1998)Lira {\em et al.}]{lira98}
Lira, P., et~al. AJ, 115, 234 

\bibitem[Mari\v c {et al}.(1977)Mari\v c {\em et al.}]{maric77}
Mari\v c, Z., Moles, M., and Vigier, J.~P. 1977, Lett.\ Nuovo Cimento, 18, 269

\bibitem[Metzger {et~al.}(1997)Metzger {\em et~al.}]{metzger97}
Metzger, M. R., et~al. 1997, Nature 387, 876

\bibitem[Narlikar(1977)Narlikar]{narlikar77}
Narlikar, J. V. 1977, Ann.\ Phys., 107, 325

\bibitem[Narlikar \& Arp(1993)Narlikar \& Arp]{na93}
Narlikar, J. V., \& Arp, H.C. 1993, ApJ., 405, 51

\bibitem[Narlikar \& Arp(1997)Narlikar \& Arp]{na97}
Narlikar, J. V., \& Arp, H.C. 1993, ApJ, 482, L119


\bibitem[Norris {et~al.}(1995)Norris {\em et~al.}]{norris95}
Norris J. P., et~al. 1995, ApJ, 439, 542

\bibitem[Nugent(1997)Nugent]{nuge_phd}
Nugent, P., 1997, Ph.~D. Thesis, University of Oklahoma, Ch. 4

\bibitem[Nugent {et~al.}(1997a)Nugent {\em et~al.}]{nuge_hyd97}
Nugent, P., Baron E., Branch D., Fisher A. \&  Hauschildt P.
1997a, ApJ., 485, 812

\bibitem[Nugent, Kim \& Perlmutter(2001)Nugent, Kim and Perlmutter]{nuge_kcorr99}
Nugent, P., Kim, A.~G. \& Perlmutter, S., 2001, PASP, submitted


\bibitem[Perlmutter {et~al.}(1995)Perlmutter {\em et~al.}]{saulaigua}
Perlmutter, S., et~al. 1995,
in Presentations at the NATO ASI in Aiguablava, Spain, LBL-38400,
  page I.1; also published in Thermonuclear Supernova, P.~Ruiz-Lapuente,
  R.~Canal, and J.Isern, editors, Dordrecht: Kluwer, page 749 (1997) 


\bibitem[Perlmutter {et~al.}(1997)Perlmutter {\em et~al.}]{perl97}
Perlmutter, S., et~al. 1997, ApJ, 483, 565

\bibitem[Perlmutter {et~al.}(1999)Perlmutter {\em et~al.}]{perl99}
Perlmutter, S., et~al., ApJ, 517, 565

\bibitem[Phillips(1993)Phillips]{phillips93}
Phillips,  M.~M., ApJ Lett., 413, L105

\bibitem[Pinto \& Eastman(2000)Pinto \& Eastman]{pi2000}
Pinto P. A. \& Eastman R. G. 2000, ApJ, 530, 744

\bibitem[Piran(1992)Piran]{piran92}
Piran T. 1992, ApJ Lett.\ 389, L45

\bibitem[Pskovskii(1977)Pskovskii]{pskovskii77}
Pskovskii, Y.P. 1977, Soviet Astron., 21, 675

\bibitem[Pskovskii(1984)Pskovskii]{pskovskii84}
Pskovskii, Y.P. 1984, Soviet Astron., 28, 658



\bibitem[Riess, Press, \& Kirshner(1995)Riess, Press, and Kirshner]{rpk95}
Riess, A.~G., Press, W.~H., \& Kirshner, R.~P. 1995, ApJ, 438, L17

\bibitem[Riess, Press, \& Kirshner(1996)Riess, Press, and Kirshner]{rpk96}
Riess, A.~G., Press, W.~H., \& Kirshner, R.~P. 1996, ApJ, 473, 88

\bibitem[Riess {et~al.}(1997)Riess {\em et~al.}]{riess_age97}
Riess, A., et~al. 1997a, AJ, 114, 722


\bibitem[Riess {et~al.}(1998)Riess {\em et~al.}]{riess98}
Riess, A., et~al. 1998, AJ, 116, 1009

\bibitem[Riess {et~al.}(1999)Riess {\em et~al.}]{riess_22}
Riess, A., et~al. 1999, AJ, 117, 707

\bibitem[Riess {et~al.}(1999a)Riess {\em et~al.}]{riess_risea}
Riess, A., et~al. 1999a, AJ, 118, 2675


\bibitem[Riess {et~al.}(1999b)Riess {\em et~al.}]{riess_riseb}
Riess, A., et~al. 1999b, AJ, 118, 2668

\bibitem[Rust(1974)Rust]{R74}
Rust, B. W.,  1974 Ph.~D. Thesis, University of Illinois; ORNL Report 4953

\bibitem[Sandage \& Perelmutter(1991)Sandage \& Perelmutter]{SP91} 
Sandage, A \& Perelmutter, J. M. 1991, ApJ, 370, 455


\bibitem[Suntzeff {et~al.}(1999)Suntzeff {\em et~al.}]{suntz98}
Suntzeff, N. B., 1999, AJ, 117, 1175



\bibitem[La Violette(1986)La Violette]{laviolette86}
La Violette, P.~A. 1986, \apj, 301, 544
 
\bibitem[Weinberg (1972) Weinberg]{weinberg72}
Weinberg, Steven, {\em Gravitation and Cosmology: Principles and Applications
of the General Theory of Relativity}, (New York: John Wiley \& Sons),
p.~30


\bibitem[Wilson(1939)Wilson]{wilson39}
Wilson, O. C. 1939, \apj, 90, 634

\bibitem[Zwicky(1929)Zwicky]{zwicky29}Zwicky, F. 1929,
Proc.\ Nat.\ Acad.\ Sci., 15, 773











 
 

\end{thebibliography}

\end{document}